\documentclass[conference]{IEEEtran}
\IEEEoverridecommandlockouts\usepackage{cite}
\usepackage{amsmath,amssymb,amsfonts}
\usepackage{algorithmic}
\usepackage{graphicx}
\usepackage{textcomp}
\usepackage{xcolor}
\usepackage{multirow}
\usepackage{float}
\usepackage{booktabs}
\usepackage{subcaption}
\usepackage{multirow}
\usepackage{threeparttable}

\def\BibTeX{{\rm B\kern-.05em{\sc i\kern-.025em b}\kern-.08em
    T\kern-.1667em\lower.7ex\hbox{E}\kern-.125emX}}
\begin{document}

\makeatletter
\newcommand{\newlineauthors}{  \end{@IEEEauthorhalign}\hfill\mbox{}\par
  \mbox{}\hfill\begin{@IEEEauthorhalign}
}

\title{Modeling Data Movement Performance on Heterogeneous Architectures}

\author{  \IEEEauthorblockN{Amanda Bienz\IEEEauthorrefmark{1},
Luke N. Olson\IEEEauthorrefmark{2},
William D. Gropp\IEEEauthorrefmark{2}
and
Shelby Lockhart\IEEEauthorrefmark{2}
}
\IEEEauthorblockA{\IEEEauthorrefmark{1}
\textit{Department of Computer Science} \\
\textit{University of New Mexico}\\
Albuquerque, USA \\
bienz@unm.edu}
\IEEEauthorblockA{\IEEEauthorrefmark{2}
\textit{Department of Computer Science} \\
\textit{University of Illinois at Urbana-Champaign}\\
Champaign, USA \\
\{lukeo,wgropp,sll2\}@illinois.edu}
}

\maketitle
\thispagestyle{plain}
\pagestyle{plain}

\begin{abstract}
The cost of data movement on parallel systems varies greatly with machine architecture, job partition, and nearby jobs.  Performance models that accurately capture the cost of data movement provide a tool for analysis, allowing for communication bottlenecks to be pinpointed.   Modern heterogeneous architectures yield increased variance in data movement as there are a number of viable paths for inter-GPU communication.  In this paper, we present performance models for the various paths of inter-node communication on modern heterogeneous architectures, including the trade-off between GPUDirect communication and copying to CPUs.  Furthermore, we present a novel optimization for inter-node communication based on these models, utilizing all available CPU cores per node.  Finally, we show associated performance improvements for MPI collective operations.
\end{abstract}

\begin{IEEEkeywords}
performance modeling, GPU, data movement, CUDA-aware, GPUDirect, MPI
\end{IEEEkeywords}

\section{Introduction}

Parallel architectures are continually advancing in compute power and energy efficiency, allowing for increasingly large high performance computing (HPC) applications.  However, the performance of parallel applications often lags behind the hardware capabilities.  There is a large effort to improve the performance and scalability of parallel applications, from numerical algorithms to machine learning methods, on state-of-the-art architectures.

The parallel performance of applications varies greatly with machine architecture, job partition, and compiler.  This performance variance is largely due to data movement, from memory access to inter-process communication.  The cost of data movement is often unpredictable, varying with the memory layer and the relative locations of sending and receiving processes.  Moreover, data movement bottlenecks are amplified on heterogeneous architectures, with high flop rates on GPUs in comparison to the limited speeds of inter-GPU data movement.  Furthermore, data movement variation is also magnified on heterogeneous architectures as the number of viable communication paths between two GPUs is increased.

Performance models, such as the max-rate model~\cite{2016GrOlSa_pingpong}, can be used to analyze the cost of communication, allowing users to pinpoint application bottlenecks.  Performance model measurements, including message latency and per-byte transport rates, can be used to analyze all communication for a single compiler on one computer.  Therefore, measurements at small scales can provide an accurate analysis of costs and bottlenecks at larger scales.  In addition, these models can expose performance bugs in applications as well as in lower-level libraries, such as MPI and CUDA\@.

In this paper, we present performance models for data movement on both Summit~\cite{summit} and Lassen~\cite{lassen}.  We fit these models to both Spectrum MPI and MVAPICH2-GDR implementations. We analyze the various paths of inter-GPU communication, and present a novel strategy for communicating large amounts of data between GPUs on different nodes.  Finally, we present MPI collective case studies, which show that the performance models accurately predict optimal communication strategies.

The remainder of the paper is outlined as follows.  Section~\ref{section:background} discusses background information, describing heterogeneous architectures and standard performance models.  Section~\ref{section:benckmarkgpu} profiles the various paths of inter-GPU data movement on Summit and Lassen and presents performance modeling results for these various paths.  Data movement optimizations that utilize all available CPU cores per node are presented in Section~\ref{section:benchmarknode} while Section~\ref{section:benchmarkmult} analyzes the benefits of these optimizations when multiple messages are sent from each GPU\@.  Case studies for the \texttt{MPI\_Alltoall} and \texttt{MPI\_Alltoallv} are presented in Section~\ref{section:casestudy}.  Finally, Section~\ref{section:conclusion} presents concluding remarks and future directions.

\section{Background}\label{section:background}

While the postal model~\cite{postal} accurately captures the cost of communicating a single message between two processes, the max-rate model~\cite{2016GrOlSa_pingpong} penalizes symmetric multiprocessing nodes with injection bandwidth limits.  

The LogP~\cite{logP} and LogGP~\cite{logGP} models improve upon the postal model with measures for the idle time that often dominates methods with synchronization, such as collective operations.  The LogP model accounts for latency bounds associated with asynchrony of processes, while the LogGP model improves upon the LogP models for large messages by adding a per-byte cost.

Locality-awareness can further improve models by distinguishing between on-socket, on-node/off-socket, and inter-node messages.  Moreover, models for irregular communication, where processes communicate many messages with non-neighboring processes, require additional parameters for accurately capturing queue search costs and estimating network contention penalties~\cite{BienzEuroMPI}.

Improved architecture-aware performance models, such as the max-rate and node-aware models, have led to the development of methods for improving communication costs.  For instance, the large performance differences between intra- and inter-node communication motivate node-aware communication optimizations on previous generation architectures~\cite{Bienz_napspmv, Bienz_napamg, Bienz_napallreduce}.

Current large-scale supercomputers, such as Summit at Oak Ridge National Laboratory and Lassen at Lawrence Livermore National Laboratory, are comprised of heterogeneous nodes, similar to Figure~\ref{fig:heteronode}.
\begin{figure}[ht!]
    \centering
    \includegraphics[width=0.7\linewidth]{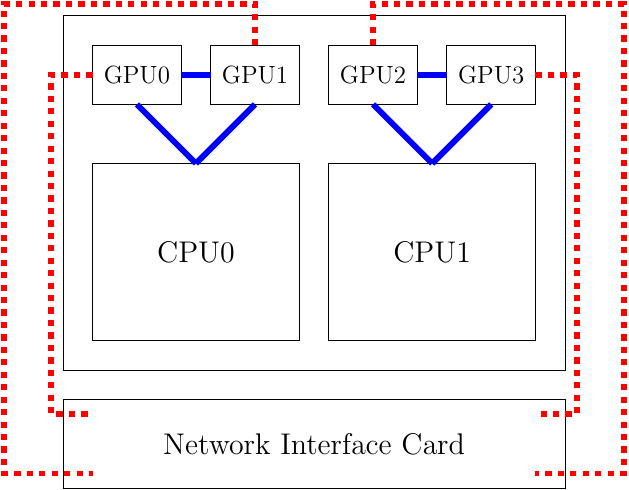}
    \caption{A schematic of a heterogeneous node.}\label{fig:heteronode}
\end{figure}
Each node of Summit, for example, contains 6 GPUs whereas each node on Lassen contains 4; both computers have $40$ CPU cores per node.  Furthermore, each node has two IBM Power9 processors with half of the GPUs directly connected to each.  Finally, links directly connect GPUs on the same Power9 processor in addition to the Network Interface Card (NIC).
\begin{figure*}[ht!]
    \centering
    \begin{subfigure}{0.4\textwidth}
        \centering
        \includegraphics[width=\textwidth]{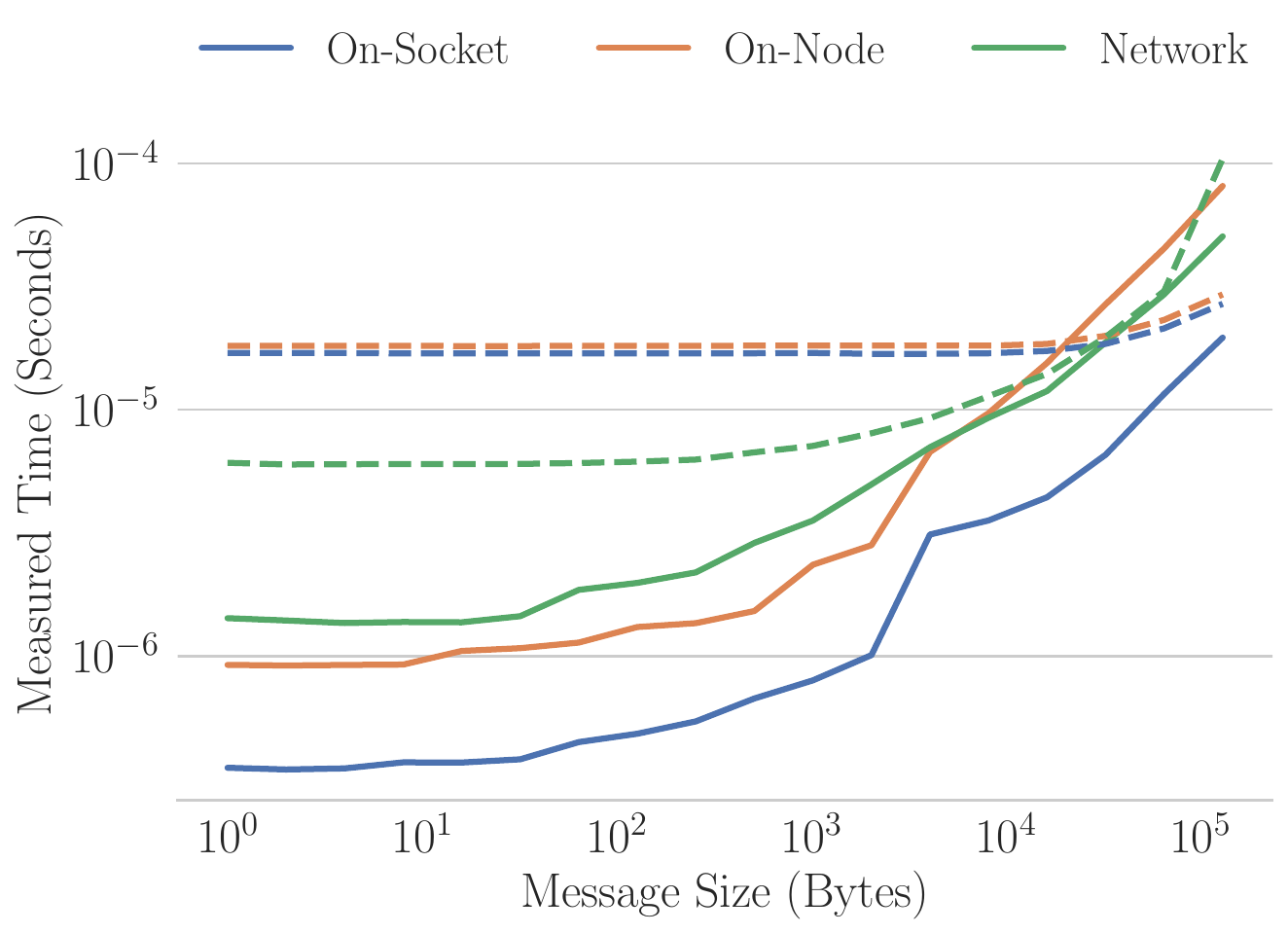}
        \caption{Summit, Spectrum MPI}\label{fig:summit_ping_pong}
    \end{subfigure}
    \hspace{2cm}
    \begin{subfigure}{0.4\textwidth}
        \centering
        \includegraphics[width=\linewidth]{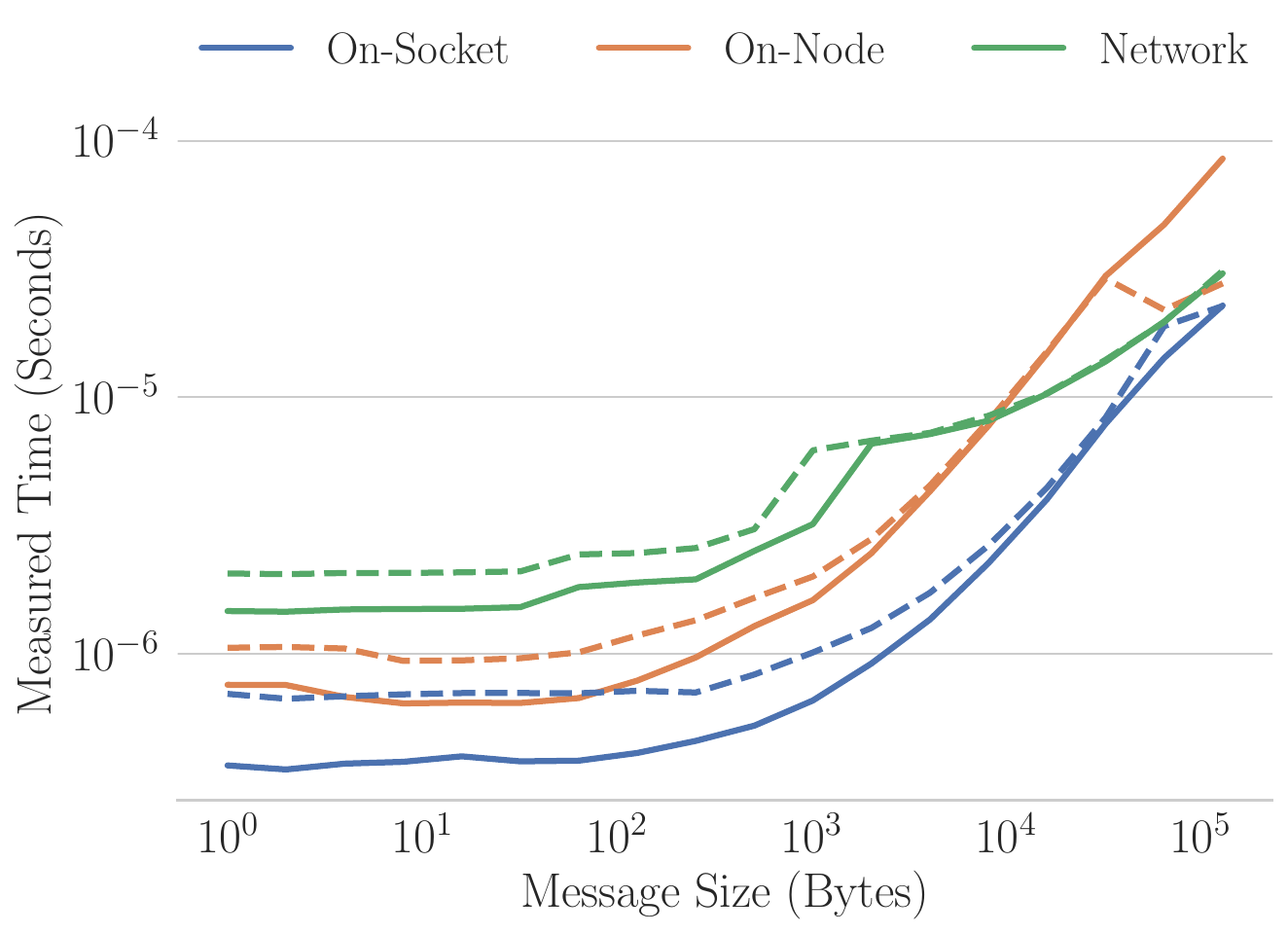}
        \caption{Lassen, MVAPICH2-GDR}\label{fig:lassen_ping_pong}
    \end{subfigure}
    \caption{The cost of communicating data directly between two CPUs (solid lines) or between two GPUs with GPUDirect (dotted lines).}\label{fig:ping_pong}
\end{figure*}

The heterogeneous architecture provides many different levels of data movement between GPUs.  For instance, two GPUs that are directly connected~---~e.g., GPU0 and GPU1 in Figure~\ref{fig:heteronode}~---~can pass messages directly.  Furthermore, inter-node data can be copied directly to the NIC without being first copied to a CPU\@.  Finally, any message can be communicated by first copying data to a CPU core.

CUDA-aware MPI allows data to be communicated between GPU memories with the MPI API\@.  GPUDirect~\cite{gpudirect} allows for data to be communicated directly between GPUs without first copying to the CPU\@.  Together, these optimizations provide increased performance of inter-GPU data movement by avoiding unnecessary copies.  Unified memory also affects data movement performance, allowing allocated memory to be accessed by both CPU and GPU cores.  These various optimizations increase the variability of data movement performance on heterogeneous architectures.

Benchmarking and performance modeling are critical in pinpointing performance bottlenecks in methods and applications.  A large variety of benchmarks are available to capture the performance of MPI applications.  However, few capture inter-GPU data movement performance. Recently, a variety of benchmarks have been extended to capture the performance of heterogeneous architectures~\cite{mpiblib, mpicuda_bench, gpubench}.

\section{Benchmarking GPU to GPU Communication}\label{section:benckmarkgpu}

Data movement bottlenecks reduce the performance and scalability of parallel applications.  Heterogeneous architectures typically have data split across a large number of GPUs; consequently, rather than relying on inter-CPU communication, data must be communicated between GPUs.  Communication bottlenecks can be determined by benchmarking and modeling the performance of inter-GPU communication.  There are two main paths of inter-GPU communication on modern heterogeneous architectures:
\begin{itemize}
    \item \textbf{CUDA-Aware GPUDirect Communication:} data is sent directly between the GPU and the NIC without being copied to the CPU, and
    \item \textbf{3-Step Communication:} data is first copied to the CPU, then communicated between two CPUs, and finally received data is copied to the destination GPU\@.
\end{itemize}

Throughout the remainder of this section, we will benchmark and model the performance of communicating a single message between two GPUs, through each viable path of communication using Spectrum MPI on Summit and MVAPICH2-GDR on Lassen.  The postal model~\cite{postal} accurately captures the cost of communicating a single message between two processes and is formulated as
\begin{equation}
    T = \alpha + \beta \cdot s \label{eqn:postal}
\end{equation}
where $\alpha$ is the message start-up cost, or latency, $\beta$ is the per-byte transport cost, and $s$ is the number of bytes to be communicated.

Modern heterogeneous architectures support CUDA-aware MPI and GPUDirect, allowing inter-node data to be transported directly to the NIC without first being copied to the CPU\@.  This path of data movement can be measured and modeled with simple postal models corresponding to CUDA-aware communication.  Each message protocol is modeled, including short, in which messages fit in the envelope and are communicated directly; eager, which assumes sufficient buffer space and communicates immediately; and finally rendezvous, which waits for the receiving process to allocate required buffer space before sending data.  The corresponding values for $\alpha$ and $\beta$ are measured for each applicable protocol.
\begin{figure*}[ht!]
    \centering
    \begin{subfigure}{0.4\textwidth}
        \centering
        \includegraphics[width=\textwidth]{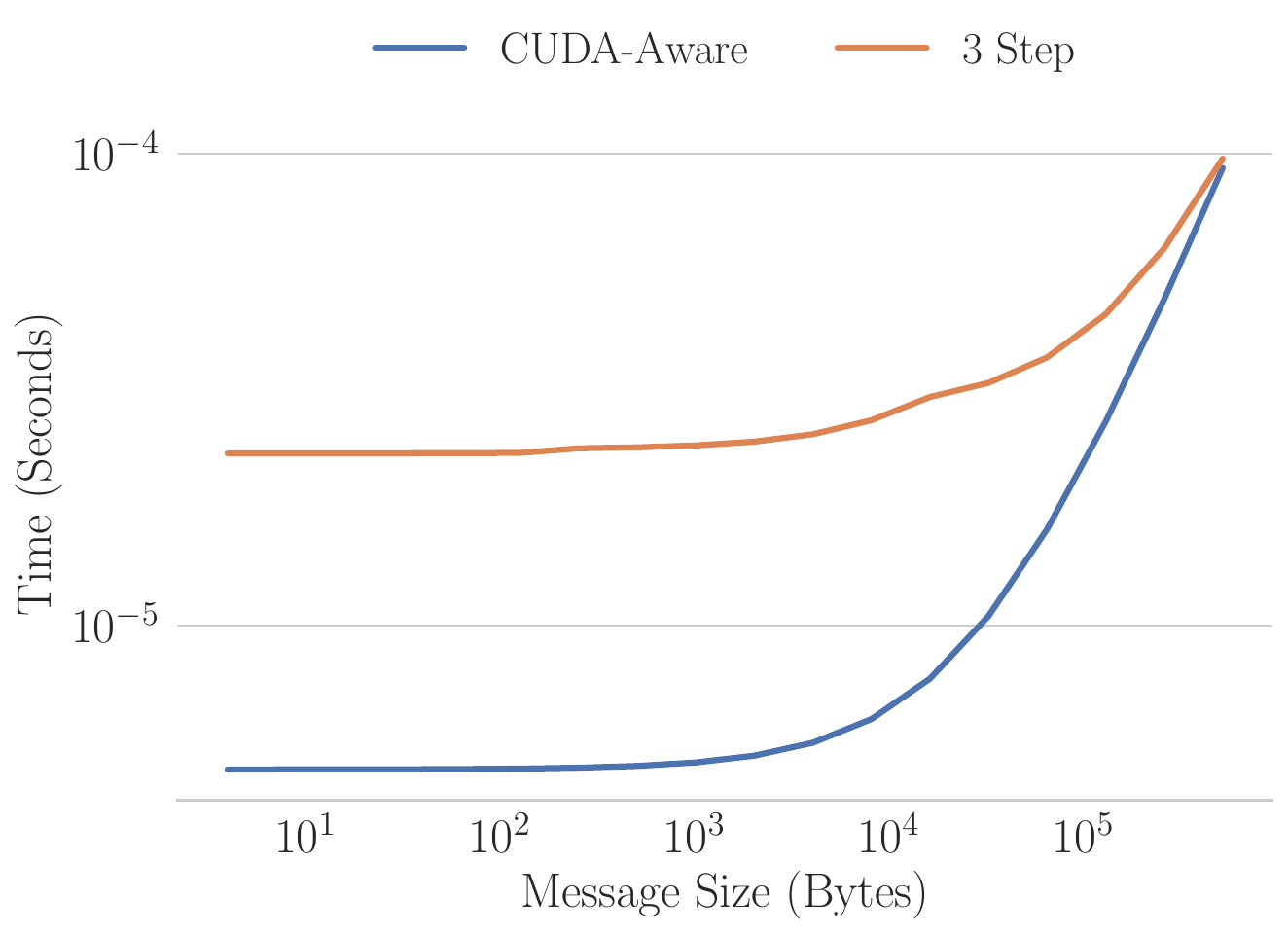}
        \caption{Summit, Spectrum MPI}\label{fig:summit_3step}
    \end{subfigure}
    \hspace{2cm}
    \begin{subfigure}{0.4\textwidth}
        \centering
        \includegraphics[width=\linewidth]{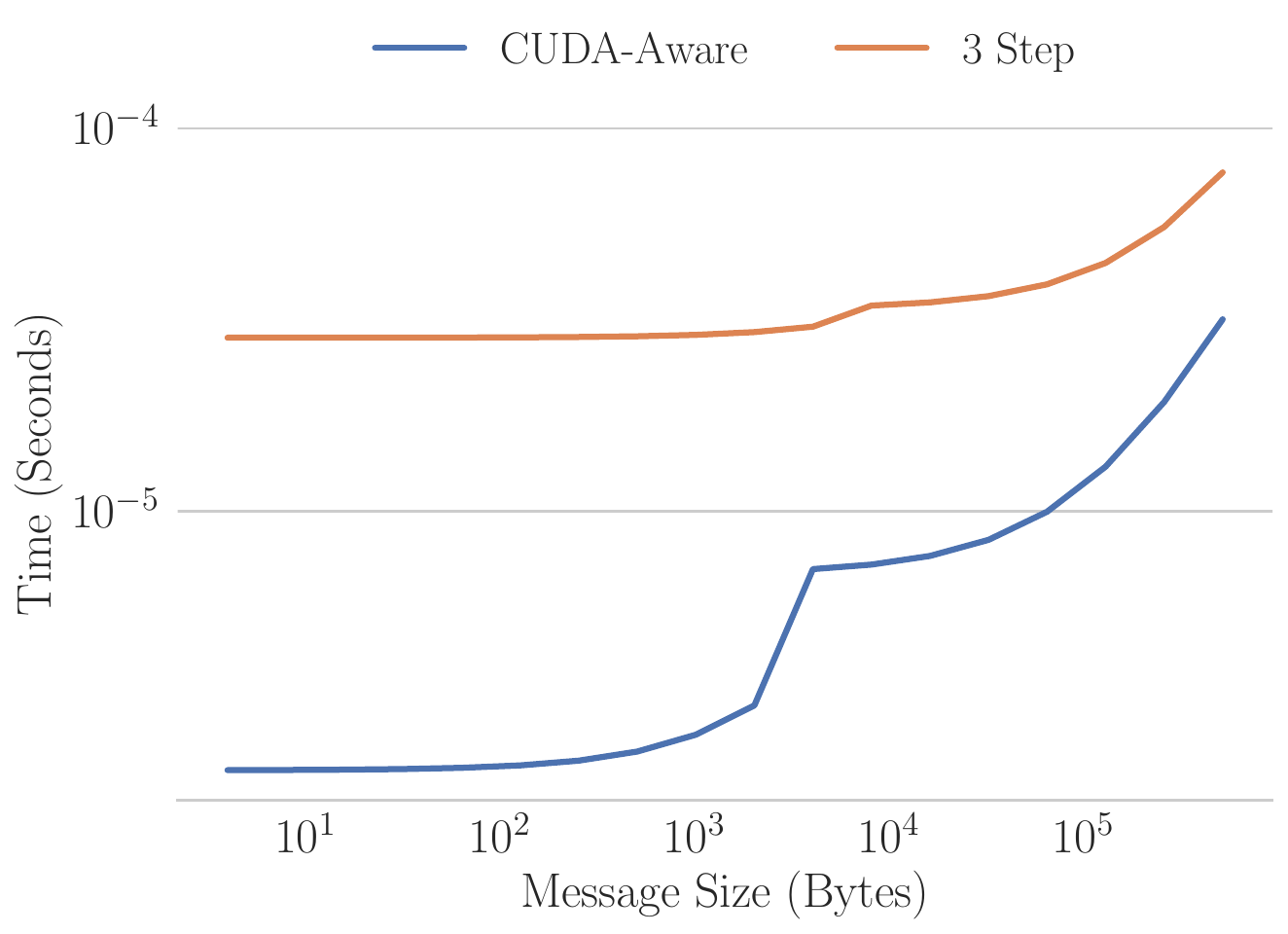}
        \caption{Lassen, MVAPICH2-GDR}\label{fig:lassen_3step}
    \end{subfigure}
    \caption{The modeled cost of inter-GPU communication for various message sizes.  Both viable paths of data movement are modeled, indicating that GPUDirect communication outperforms the 3-step copy to CPU method when communicating a single message between GPUs.}\label{fig:inter_gpu}
\end{figure*}

Figure~\ref{fig:ping_pong} highlights the cost of transporting data between a set of CPUs (solid lines) or GPUs with GPUDirect (dotted lines).  The communication costs are split into on-socket, for which communicating CPU cores lie on the same NUMA node and GPUs are directly connected via a link; on-node, for which data must cross NUMA node regions and GPUs are not connected by a direct link; and network messages, which are transported directly to the NIC before being communicated across the interconnect.  The parameters associated with the postal models for both inter-CPU and inter-GPU communication are displayed in Table~\ref{table:postal}.
\begin{table}[ht!]
    \renewcommand{\arraystretch}{1.1}
    \centering
    \begin{threeparttable}
    \begin{tabular}{p{1mm}p{1mm}ccr|ccc}
      \toprule
        & & & & & on-socket & on-node & off-node \\
        \midrule 
        \parbox[t]{1mm}{\multirow{8}{*}{\rotatebox[origin=c]{90}{Summit}}}
        & \parbox[t]{1mm}{\multirow{8}{*}{\rotatebox[origin=c]{90}{(Spectrum MPI)}}}
        & \multirow{6}{*}{CPU}
        & \multirow{2}{*}{Short} 
        & $\alpha$\tnote{*} & 3.51e-07 & 9.08e-07 & 1.38e-06 \\
        & & & & $\beta$\tnote{\textdagger} & 2.62e-10 & 1.46e-09 & 3.82e-10 \\
        \cline{6-8}
        & & & \multirow{2}{*}{Eager} 
        & $\alpha$\tnote{*} & 4.73e-07 & 1.17e-06 & 1.85e-06 \\
        & & & & $\beta$\tnote{\textdagger} & 6.95e-11 & 2.16e-10 & 2.93e-10 \\
        \cline{6-8}
        & & & \multirow{2}{*}{Rend} 
        & $\alpha$\tnote{*} & 2.46e-06  & 5.81e-06 & 6.56e-06 \\
        & & & & $\beta$\tnote{\textdagger} & 3.31e-11 & 1.46e-10 & 8.51e-11 \\
        \cline{3-8}
        & & \multirow{2}{*}{GPU}
        & \multirow{2}{*}{} 
        & $\alpha$\tnote{*} & 1.68e-05 & 1.80e-05 & 4.96e-06 \\
        & & & & $\beta$\tnote{\textdagger} & 1.86e-11 & 2.09e-11 & 1.69e-10 \\

        \midrule 
        \parbox[t]{1mm}{\multirow{8}{*}{\rotatebox[origin=c]{90}{Lassen}}} 
        & \parbox[t]{1mm}{\multirow{8}{*}{\rotatebox[origin=c]{90}{ (MVAPICH2-GDR)}}}
        & \multirow{4}{*}{CPU}
        & \multirow{2}{*}{Eager} 
        & $\alpha$\tnote{*} & 3.99e-07 & 7.07e-07 & 1.53e-06 \\
        & & & & $\beta$\tnote{\textdagger} & 5.59e-11 & 2.23e-10 & 4.38e-10 \\
        \cline{6-8}
        & & & \multirow{2}{*}{Rend} 
        & $\alpha$\tnote{*} & 3.62e-06 & 1.07e-05 & 6.90e-06 \\
        & & & & $\beta$\tnote{\textdagger} & 3.71e-11 & 1.42e-10 & 4.63e-11 \\
        \cline{3-8}
        & & \multirow{4}{*}{GPU}
        & \multirow{2}{*}{Eager} 
        & $\alpha$\tnote{*} & 7.09e-07 & 1.04e-06 & 2.11e-06 \\
        & & & & $\beta$\tnote{\textdagger} & 5.79e-11 & 2.15e-10 & 4.91e-10 \\
        \cline{6-8}
        & & & \multirow{2}{*}{Rend} 
        & $\alpha$\tnote{*} & 6.39e-06 & 2.61e-05 & 6.87e-06 \\
        & & & & $\beta$\tnote{\textdagger} & 3.38e-11 & 4.59e-13 & 4.73e-11 \\
     \bottomrule
    \end{tabular}
    * measured in seconds \hspace{1cm} \textdagger measured in seconds per byte
                        \caption{Measured parameters for inter-CPU and inter-GPU communication. Note that messaging protocol delineation for inter-GPU communication on Summit has been excluded due to an insignificant difference in performance between protocols.}\label{table:postal}
    \end{threeparttable}
\end{table}

A 3-step communication strategy requires copying all data from a GPU to a corresponding CPU, communicating between CPU cores, and finally copying data to the destination GPU\@.  Corresponding models require profiling not only the cost of communicating between sets of CPUs, but also copying data between the CPUs and GPUs. 
\begin{table}[ht!]
    \centering
    \renewcommand{\arraystretch}{1.1}
    \begin{threeparttable}
    \begin{tabular}{rrc|cc}
      \toprule
      & & & HostToDevice & DeviceToHost \\
      \midrule
        \parbox[t]{2mm}{\multirow{4}{*}{\rotatebox[origin=c]{90}{Summit}}} 
        &\multirow{2}{*}{on-socket} & $\alpha$\tnote{*} & 1.09e-05 & 1.09e-05\\
        & & $\beta$\tnote{\textdagger} & 2.38e-11 & 2.36e-11 \\
        \cline{2-5}
        & \multirow{2}{*}{off-socket} & $\alpha$\tnote{*} & 1.26e-05 & 1.25e-05\\
        & & $\beta$\tnote{\textdagger} & 2.71e-11 & 2.72e-11\\
        \midrule
        \parbox[t]{2mm}{\multirow{4}{*}{\rotatebox[origin=c]{90}{Lassen}}} 
        &\multirow{2}{*}{on-socket} & $\alpha$\tnote{*} & 1.33e-05 & 1.35e-05\\
        & & $\beta$\tnote{\textdagger} & 1.80e-11 & 1.75e-11 \\
        \cline{2-5}
        & \multirow{2}{*}{off-socket} & $\alpha$\tnote{*} & 1.42e-05 & 1.40e-05\\
        & & $\beta$\tnote{\textdagger} & 2.84e-11 & 2.83e-11\\
        \bottomrule
    \end{tabular}
                    * measured in seconds \hspace{7mm} \textdagger measured in seconds per byte
        \caption{Measured parameters for \texttt{cudaMemcpyAsync}.}\label{table:memcpy}
    \end{threeparttable}
\end{table}
Table~\ref{table:memcpy} displays the postal model parameters associated with performing \texttt{cudaMemcpyAsync} between a GPU and CPU, both for transferring from the host to the device as well as from the device to the host.  The cost of both on- and off-socket data transfers are presented, with on-socket data transfers copying between a CPU and GPU that are directly connected, while off-socket data transfers require traversing non-uniform memory access (NUMA) regions.

The model parameters presented in Tables~\ref{table:postal} and ~\ref{table:memcpy} can be used to analyze the cost of the various paths of inter-node data movement.
Figure~\ref{fig:inter_gpu} displays the modeled costs of communicating inter-node messages of various sizes between two GPUs through either GPUDirect communication or 3-step communication, which requires two on-socket \texttt{cudaMemcpyAsync} operations.  These models indicate that when sending a single message between a set of GPUs, GPUDirect is more efficient for all modeled sizes.  The difference is most drastic for small messages due to the large latency associated with \texttt{cudaMemcpyAsync}.

\section{Utilizing Many CPU Processes}\label{section:benchmarknode}

Modern heterogeneous nodes are comprised of multiple GPUs and also contain many CPU cores.  The max-rate model
\begin{equation}\label{eqn:maxrate}
    T = \alpha + \frac{\texttt{ppn} \cdot s}{\max{(R_{N}, R_{p} \cdot \texttt{ppn})}}
    \beta \cdot s
\end{equation}
captures the cost of communication when multiple processes are active per node.  In this model, $\texttt{ppn}$ is equal to the number of communicating processes per node, $R_{p}$ is the inter-process data transport rate, and $R_{N}$ is the rate at which data can be injected into the network.  If $R_{p} \cdot \texttt{ppn}$ is less than $R_{N}$, this model reduces to the postal model from Equation~\ref{eqn:postal}.
\begin{table}[ht!]
    \centering
    \begin{tabular}{rr|c}
      \toprule
     & & $R_{N}$ \\
     & & [bytes/sec]\\
     \midrule
    \multirow{2}{*}{Summit}
    & inter-CPU &  3.0e-11\\
    & inter-GPU &  5.1e-11\\
    \midrule
    \multirow{2}{*}{Lassen}
    & inter-CPU &  2.5e-11\\
    & inter-GPU &  --\\
    \bottomrule
    \end{tabular}
    \medskip
    \caption{Measured parameters for injection bandwidth limits.  Note, inter-GPU injection bandwidth limit are excluded for Lassen, as these limits were not reached with the four available GPUs per node.}\label{table:maxrate}
\end{table}
Table~\ref{table:maxrate} displays the injection bandwidth rate for both inter-CPU and inter-GPU communication.

\begin{figure}[ht!]
    \centering
    \includegraphics[width=.8\linewidth]{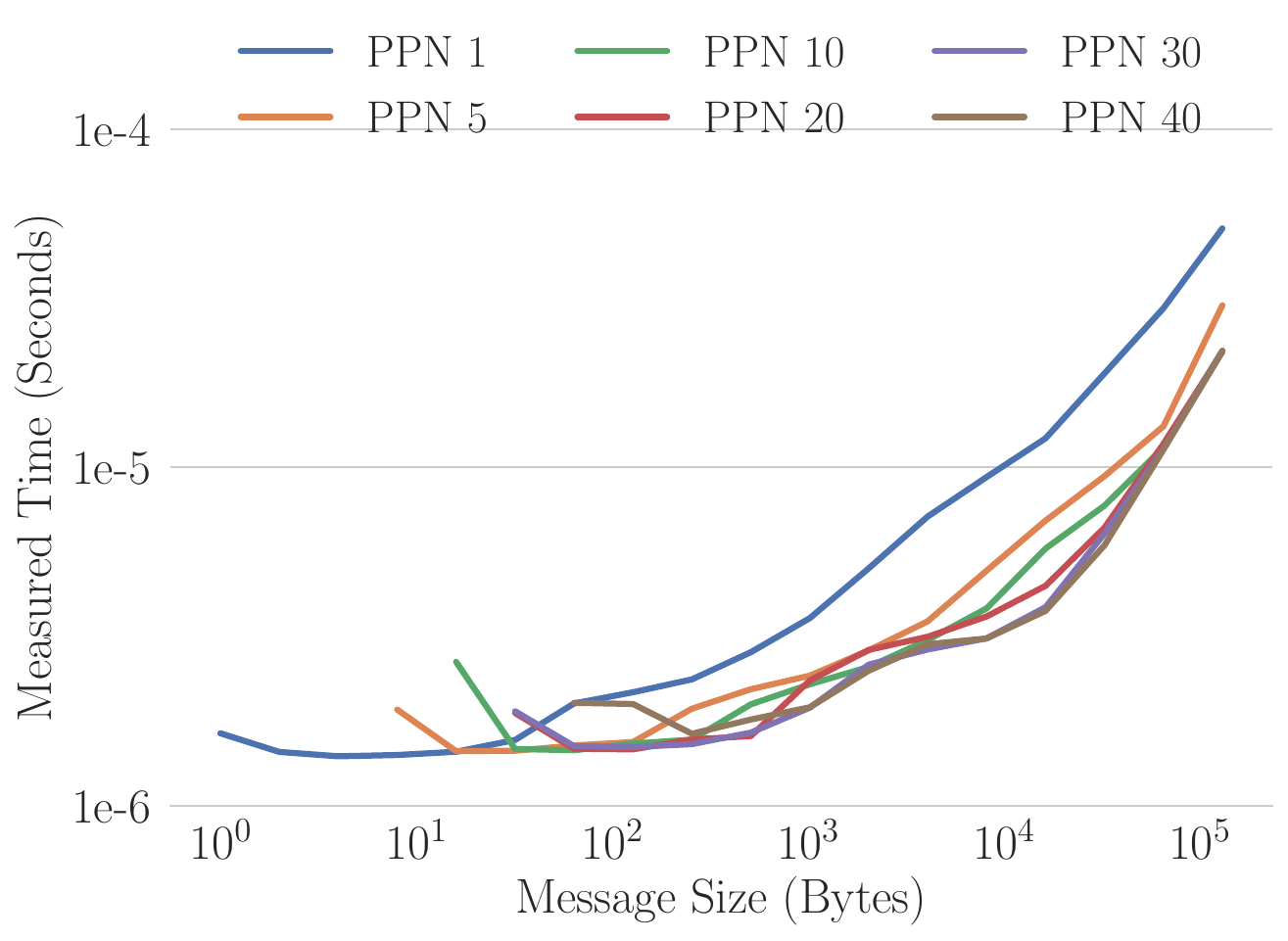}
    \caption{The cost of communicating data between two nodes of Summit, splitting the data evenly across PPN CPU processes.  Note, Lassen results were excluded due to similarities.}\label{fig:node_pong}
\end{figure}
\begin{figure*}[ht!]
    \centering
    \begin{subfigure}{0.4\textwidth}
        \centering
        \includegraphics[width=\textwidth]{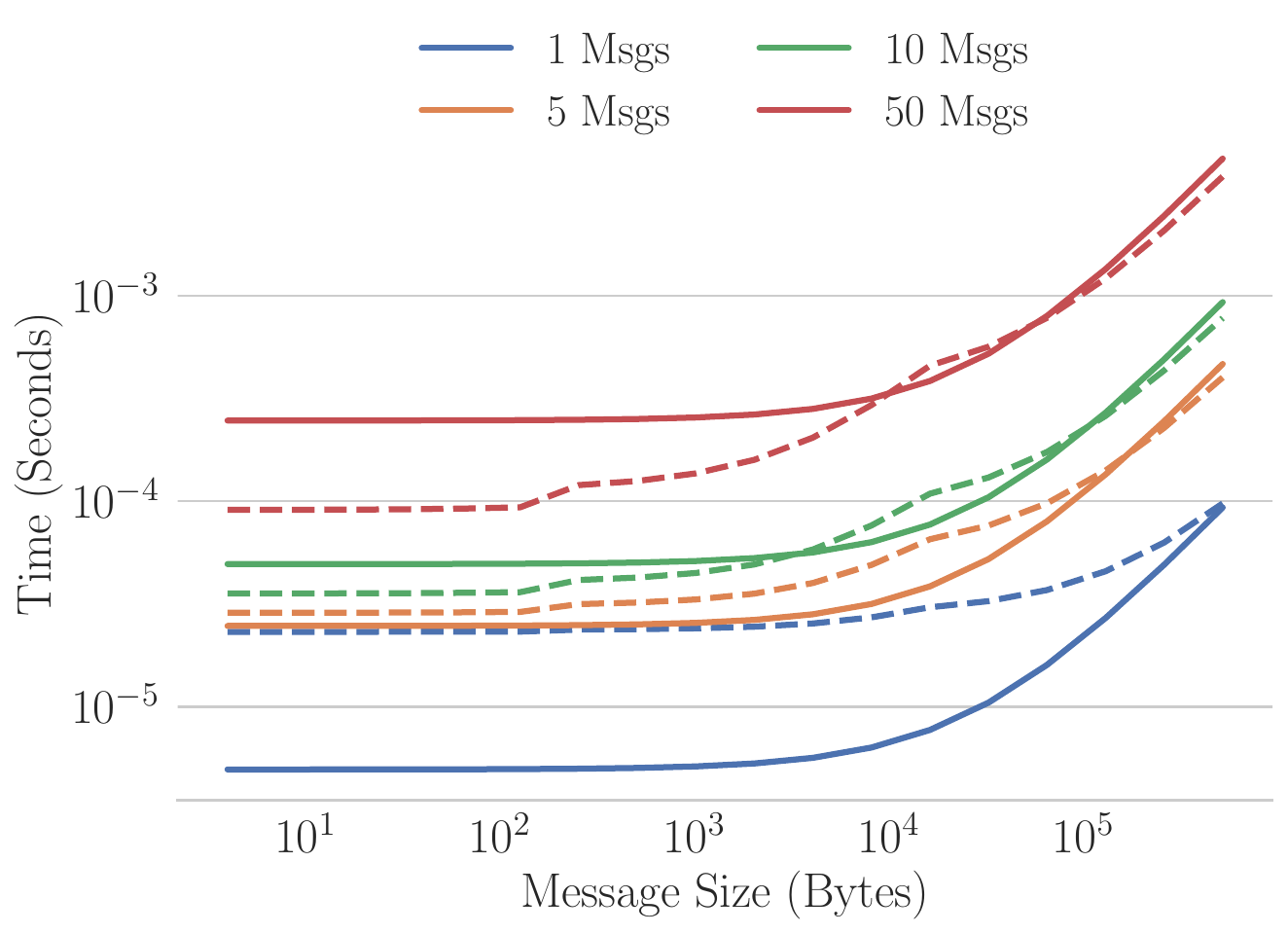}
        \caption{Summit, Spectrum MPI}\label{fig:summit_3step_mult}
    \end{subfigure}
    \hspace{2cm}
    \begin{subfigure}{0.4\textwidth}
        \centering
        \includegraphics[width=\linewidth]{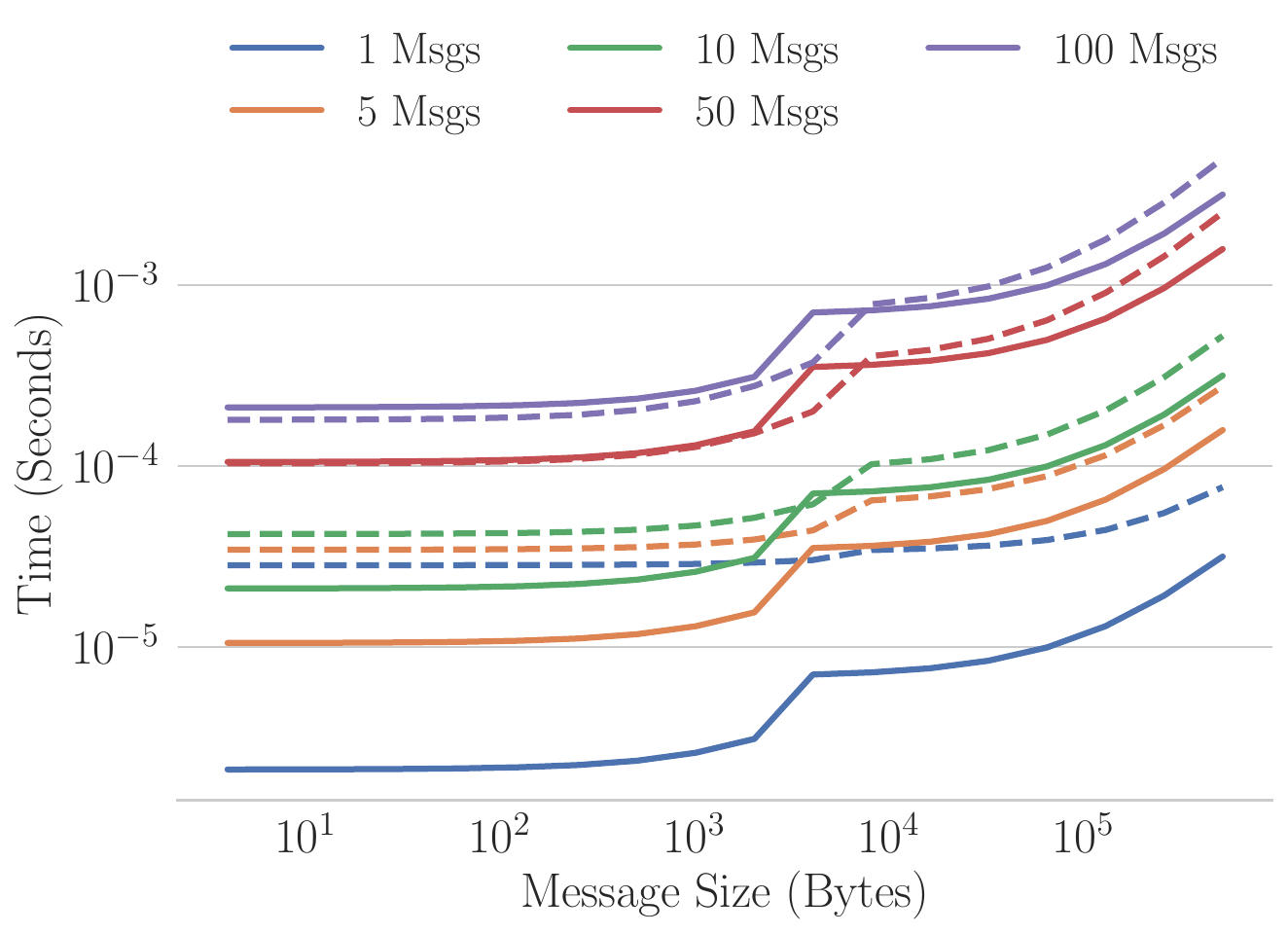}
        \caption{Lassen, MVAPICH2-GDR}\label{fig:lassen_3step_mult}
    \end{subfigure}
    \caption{The modeled cost of GPUDirect (solid) and 3-Step copy to CPU (dotted) approaches when communicating between $1$ and $100$ messages of various sizes.  
}\label{fig:three_step_mult}
\end{figure*}
Figure~\ref{fig:node_pong} displays the measured cost of communicating various amounts of data between two nodes, splitting the data across a portion of the available $40$ CPU cores per node.  Even with injection bandwidth limits, it is most efficient to have all processes active in inter-node communication, with data evenly split across them.  Therefore, the three-step inter-GPU communication can be further optimized by utilizing all available CPUs rather than copying data to a single CPU\@.

There are multiple methods for distributing data across all CPU cores per node.  Data can be copied to a single CPU core with \texttt{cudaMemcpyAsync}, and then communicated to the other processes through intra-node MPI communication.  Alternatively, CUDA Multi-Process Service (MPS) allows multiple processes to overlap \texttt{memcpy} operations.  Furthermore, an allocated region of device memory can be shared among multiple MPI processes.  Therefore, multiple CPU processes can each use \texttt{memcpy} to move a portion of the data from the GPU, enabling the data to be evenly distributed across all processes without extra MPI intra-node communication.

\section{Optimizing Multiple Messages}\label{section:benchmarkmult}

In practice, applications often require multiple messages to be sent from and received by each GPU\@.  The cost of communicating multiple messages can be modeled with the max-rate model used for each message such as
\begin{equation}\label{eqn:maxrate_mult}
    T = \alpha \cdot n + \frac{\texttt{ppn} \cdot s}{\max{(R_{N}, R_{p} \cdot \texttt{ppn})}}
    \beta \cdot s
\end{equation}
where $n$ is the number of messages communicated from any process.  As the latency term $\alpha$ is correlated with the number of messages, larger message counts greatly reduce the performance of GPUDirect communication, which has significantly higher latency than inter-CPU communication.

While the latency of inter-CPU communication is much smaller than that between GPUs, there is a significant latency associated with \texttt{cudaMemcpyAsync}.  However, this operation can be performed one time regardless of the number of messages to be communicated.  Furthermore, as the same data is often communicated in multiple messages, there is also the potential for the \texttt{cudaMemcpyAsync} to be significantly smaller than the inter-CPU communication as each data value only needs to be copied to the CPU once.

Figure~\ref{fig:three_step_mult} shows the modeled speedup associated with copying to the CPU versus GPUDirect MPI when sending various message counts.  Assuming no data is duplicated, the models indicate that copying to the CPU is faster than GPUDirect for nearly all message sizes when sending at least $10$ messages on Summit with Spectrum MPI.  However, when using MVAPICH2-GDR on Lassen, around $100$ messages are required before this three-step approach pays off.  There is the potential for additional speedup at all message counts if the same data values are sent in multiple messages, decreasing the size of the \texttt{cudaMemcpyAsync} operation.  Furthermore, speedups can be amplified by evenly splitting data across all available CPU cores.
\begin{figure*}[ht!]
    \centering
    \begin{subfigure}{\textwidth}
        \centering
        \includegraphics[width=0.4\textwidth]{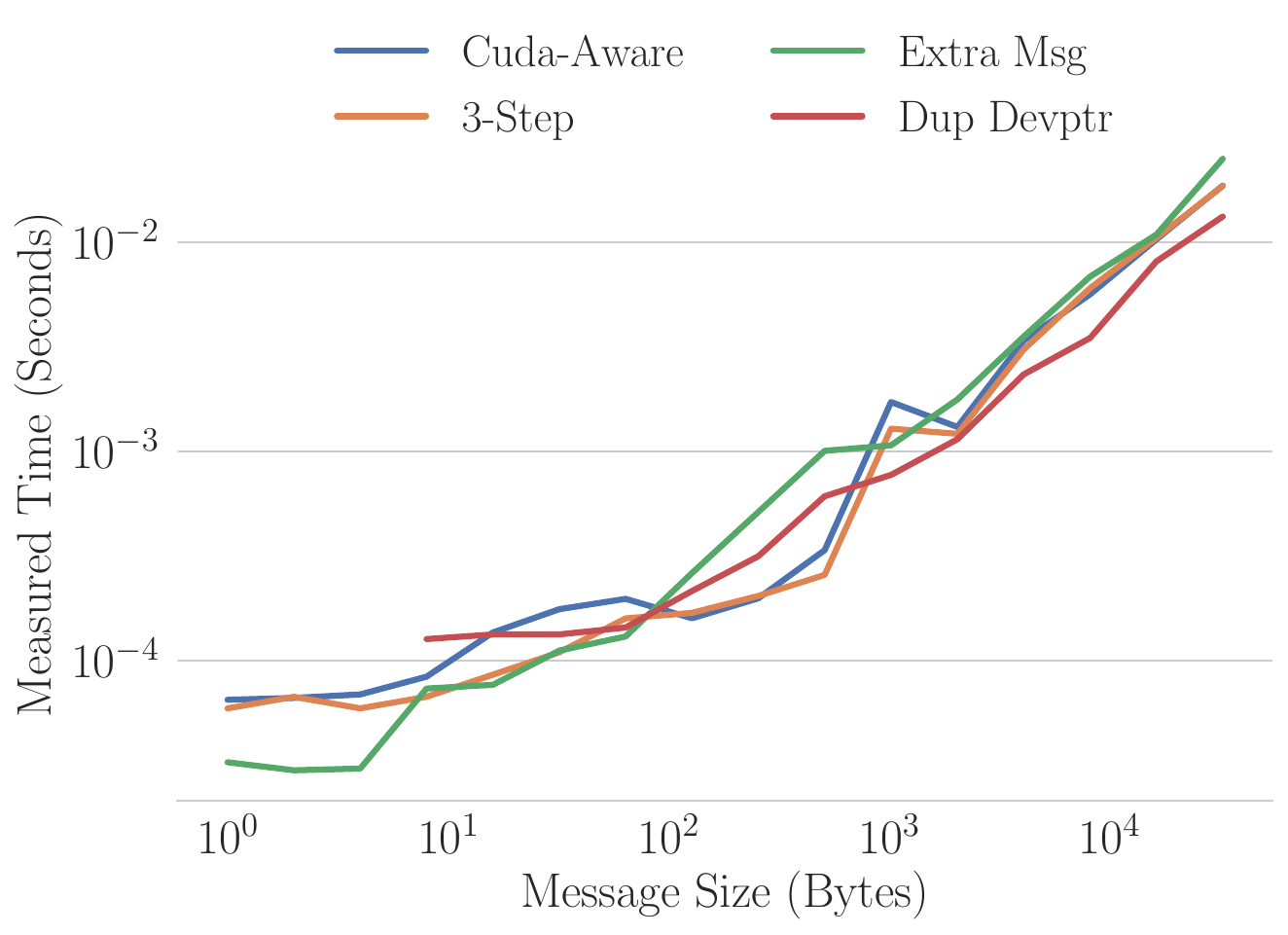}
        \hspace{2cm}
        \includegraphics[width=0.4\textwidth]{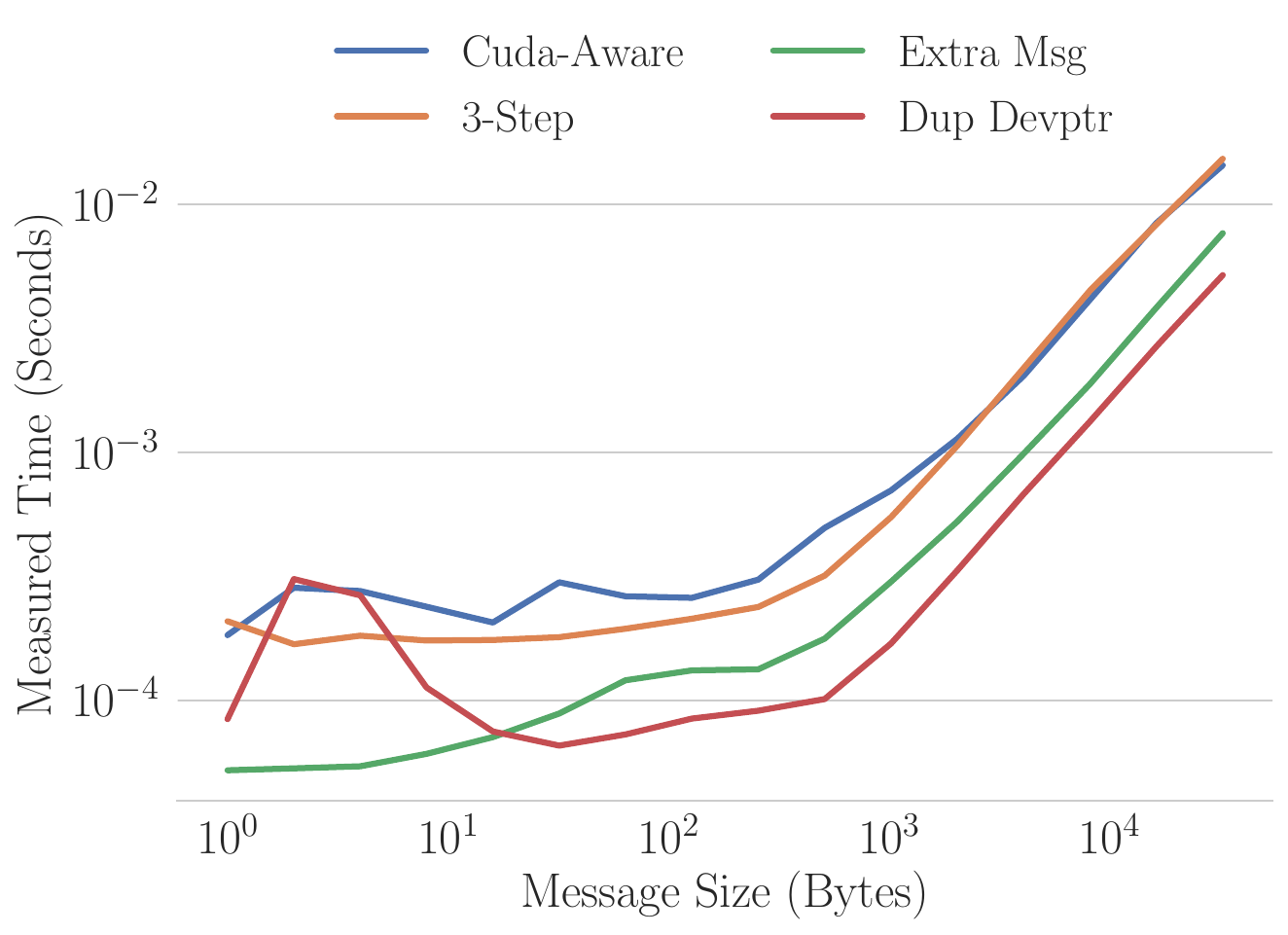}
        \caption{Summit, Spectrum MPI, \texttt{MPI\_Alltoall} (left) and \texttt{MPI\_Alltoallv} (right)}\label{fig:summit_collective}
    \end{subfigure}\\
    \begin{subfigure}{\textwidth}
        \centering
        \includegraphics[width=0.4\linewidth]{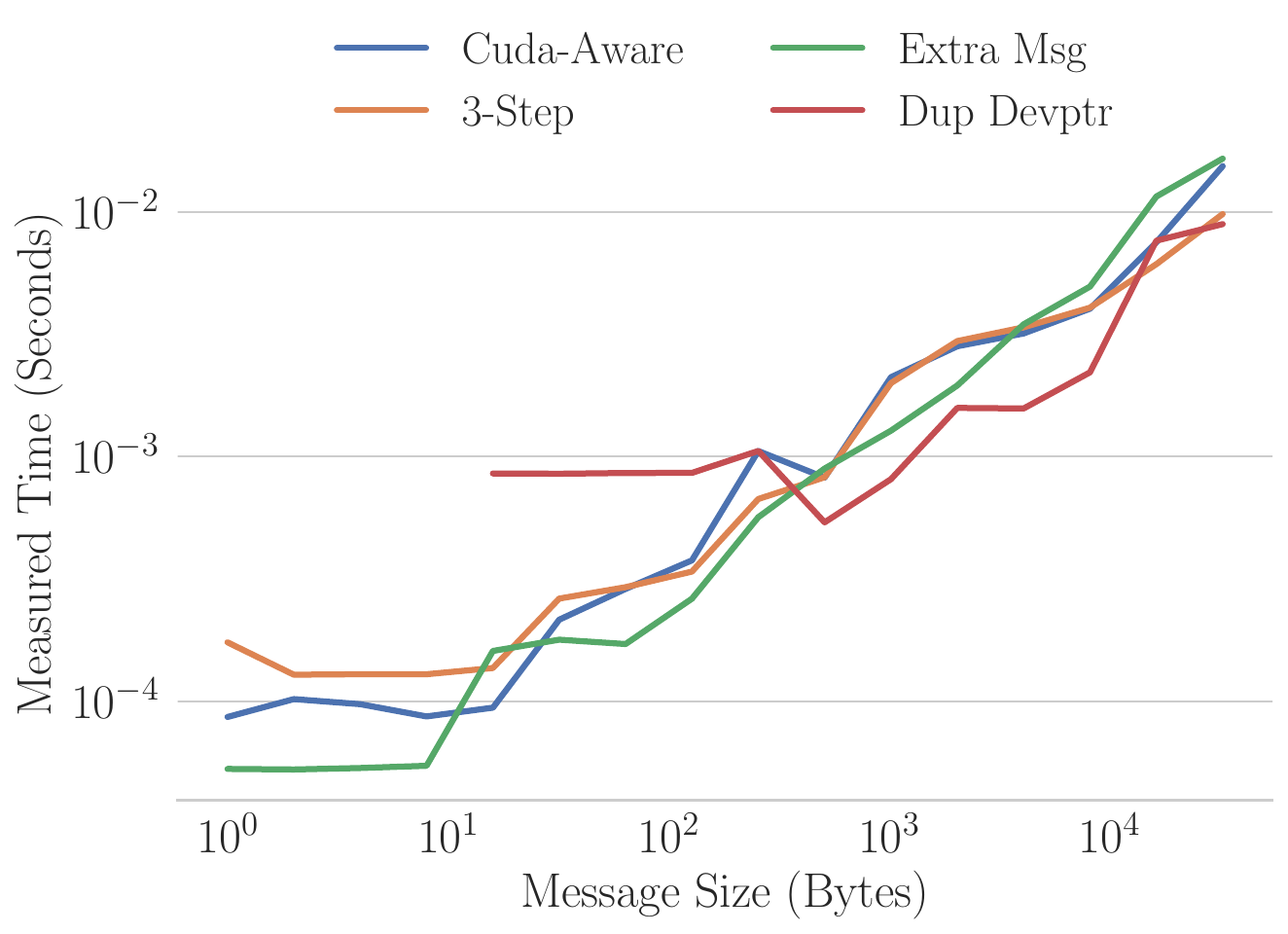}
        \hspace{2cm}
        \includegraphics[width=0.4\linewidth]{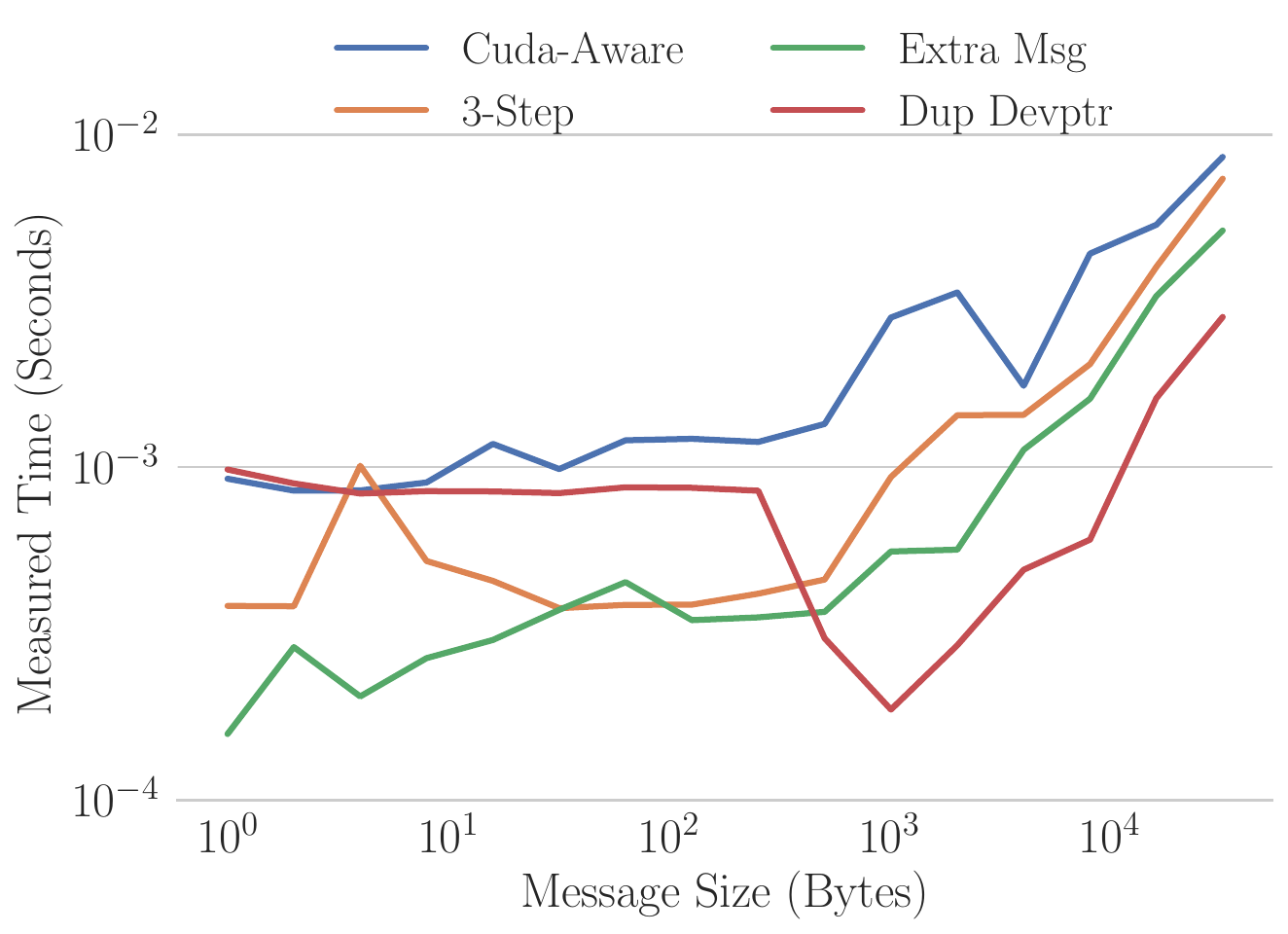}
        \caption{Lassen, MVAPICH2-GDR, \texttt{MPI\_Alltoall} (left) and \texttt{MPI\_Alltoallv} (right)}\label{fig:lassen_collective}
    \end{subfigure}
    \caption{The performance of MPI Collectives across $32$ nodes, using the various methods of communication.}\label{fig:case_studies}
\end{figure*}

\section{Case Studies: MPI Collectives}\label{section:casestudy}

MPI collective operations require communication of data among all processes in an MPI communicator.  On heterogeneous architectures, the data is typically communicated between all GPUs.  Therefore, the number of messages is proportional to the active number of GPUs.

CUDA-aware implementations of the MPI collectives utilize GPUDirect in both Spectrum MPI and MVAPICH2-GDR.  However, the performance models in Section~\ref{section:benchmarkmult} show that when communicating a large number of messages, the three-step copy to CPU approach should outperform the CUDA-aware algorithm.  Furthermore, the models show speedup for large messages sizes when data is distributed across all CPU processes.

All CPU processes are utilized during the MPI collectives by distributing data across all available cores so that each core holds a portion of the data to be sent to each GPU\@.  For example, as Summit has $6$ GPUs and $40$ CPU cores per node, $6$ CPU cores are utilized per $GPU$.  A GPU rank between $0$ and $5$ is then assigned to each CPU core.  Each of the CPU cores holds $\frac{1}{6}^{th}$ of the data and performs the collective on this smaller data size among all processes with the same GPU rank.  This optimizes inter-node communication by having each available CPU core communicate an equal portion of data.

Four different options are tested for \texttt{MPI\_Alltoall}and \texttt{MPI\_Alltoallv}:

    \noindent \textbf{1. CUDA-Aware: } data allocated in GPU memory is passed to the collective operation, utilizing GPUDirect to avoid copying data to the CPUs.
    
    \noindent \textbf{2. 3-Step: } all data is copied to a single CPU. An inter-CPU collective is then performed, and finally all received data is copied from the receiving CPU to the destination GPU.
    
    \noindent \textbf{3. Extra Msg: } all data is copied to a single CPU.  Data is then redistributed across all available CPU cores per GPU so that each process holds an equal portion of the values to be sent to each GPU\@.  Each process then calls the collective operation on this smaller portion of data.  Received data is then sent back to a single CPU core per GPU\@.  Finally, a single CPU transfers all received data to the destination GPU.
    
    \noindent \textbf{4. Dup Devptr: } each CPU core per GPU transfers a portion of the data from the GPU before calling the collective on this portion of data. Each process then transfers the portion of received data back to the GPU\@.
    
All communication between sets of CPUs or sets of GPUs is performed with MPI, which data is moved between a CPU and GPU with \texttt{cudaMemcpyAsync}.  A single test consists of performing a collective many times to reach timer precision and taking the maximum time across all processes. Furthermore, each test is performed three times, and the minimum of all timings is presented.  The methods are tested using all available GPUs on 32 nodes, using Spectrum MPI on Summit and MVAPICH2-GDR on Lassen.

The \texttt{MPI\_Alltoall} operation consists of distributing equal amounts of data to every other process, yielding a large number of messages.  When distributing data across all available CPU cores, each core sends an equal portion of the data.  However, in the tested implementation, utilizing all CPU cores does not reduce the number of messages per process.

The \texttt{MPI\_Alltoallv} consists of sending messages of any size to each of the other processes, allowing for communication among only a subset of the processes.  Both Spectrum MPI and MVAPICH-GDR implementations are outperformed in many cases by using \texttt{MPI\_Isend} and \texttt{MPI\_Irecv} for each message, and waiting for all messages to complete.  Therefore, this study tests both the existing implementation of \texttt{MPI\_Alltoallv} as well as communicating with point-to-point messages, displaying the least costly of these two strategies.  When distributing the point-to-point communication across CPU cores, the per-core message count is reduced.  For example, on Summit, each of the $6$ CPU cores per GPU sends $\frac{1}{6}^{\textnormal{th}}$ of the messages.

Figure~\ref{fig:case_studies} shows the cost of performing \texttt{MPI\_Alltoall} and \texttt{MPI\_Alltoallv} with each of the communication strategies.  Equal-sized messages are sent to each GPU.  In all cases, the extra message approach outperforms all others for very small messages and duplicate device pointer performs best for very large messages.  To achieve greater speedups, the \texttt{MPI\_Alltoall} can be implemented equivalently to the \texttt{MPI\_Alltoallv}.  On Lassen, there is a large overhead associated with duplicate device pointers for very small messages.  This overhead is consistently seen across all versions of MPI on Lassen, and is not related to MVAPICH2-GDR.

\section{Conclusion and Future Work}\label{section:conclusion}

Accurate performance models are a useful tool that guide the development of parallel applications.  Performance models for the various paths of inter-GPU communication show that, while GPUDirect communication is optimal when sending a single small message between GPUs, programs that send a number of messages perform optimally by copying the data to the CPU\@.  Furthermore, the performance can be further improved by evenly distributing the data across all available CPU cores so that each process communicates a smaller amount of data, and in some cases fewer messages, through the network.

This work can be extended to other heterogeneous architectures and alternative MPI implementations.  Furthermore, while this paper focuses on inter-node communication, performance models for intra-node communication can also be analyzed to improve application bottlenecks.  Finally, optimizations from this paper can be applied to HPC applications, as a large number of inter-GPU messages are required for many applications from stencil codes to sparse matrix operations.

\section*{Acknowledgment}

This material is based in part upon work supported by the Department of Energy, National Nuclear Security Administration, under Award Number DE-NA0002374. 
\bibliographystyle{IEEEtran}
\bibliography{refs}

\end{document}